\documentclass{ws-rv9x6}
\usepackage{subfigure}   
\usepackage{ws-rv-thm}   
\usepackage{ws-rv-van}   
\makeindex

\begin{document}

\chapter[Can one measure the Cosmic Neutrino Background?]{Can one measure the Cosmic Neutrino Background?}\label{ra_ch1}

\author[Amand Faessler, Rastislav Hod\'ak, Sergey Kovalenko, Fedor \v Simkovic]{Amand Faessler\footnote{Author footnote.}}

\address{Institute f\"{u}r Theoretische Physik der Universit\"{a}t T\"{u}bingen,\\
D-72076 T\"{u}bingen, Germany; faessler@uni-tuebingen.de\footnote{Affiliation footnote.}}

\author[Amand Faessler, Rastislav Hod\'ak, Sergey Kovalenko, Fedor \v Simkovic]{Rastislav Hod\'ak}
\address{Institute of Experimental and Applied Physics,\\ Czech Technical University in Prague,
CZ-128 00 Prague, Czech Republic}

\author[Amand Faessler, Rastislav Hod\'ak, Sergey Kovalenko, Fedor \v Simkovic]{Sergey Kovalenko}
\address{Universidad T\'ecnica Federico Santa Mar\'\i a, Centro-Cient\'\i fico-Tecnol\'{o}gico de Valpara\'\i so, Casilla 110-V, Valpara\'\i so, Chile}

\author[Amand Faessler, Rastislav Hod\'ak, Sergey Kovalenko, Fedor \v Simkovic]{Fedor \v Simkovic}
\address{JINR, 141980 Dubna, Moscow Region, Russia;\\
Faculty of Mathematics, Physics and Informatics, Comenius University, SK-842 48 Bratislava, Slovakia}

\begin{abstract}
The Cosmic Microwave Background (CMB) yields information about our Universe at around 380 000 years after
the Big Bang (BB). Due to the weak interaction of the neutrinos with matter the Cosmic Neutrino Background
(CNB) should give information about a much earlier time of our Universe, around one second after the Big
Bang. Probably the most promising method to `see' the Cosmic Neutrino Background is the capture of the
electron neutrinos from the Background by Tritium, which then decays into $^3$He and an electron with
the energy of the the Q-value = 18.562 keV plus the electron neutrino rest mass. The `KArlsruhe TRItium
Neutrino' (KATRIN) experiment, which is in preparation, seems presently the most sensitive proposed
method for measuring the electron antineutrino mass. At the same time KATRIN can also look by the
reaction $\nu_e$($\sim$  1.95  Kelvin) + $^3$H $\rightarrow$ $^3$He + e$^-$ (Q = 18.6  keV +  m$_{\nu e}$c$^2$).
The capture of the Cosmic Background Neutrinos (CNB) should show in the electron spectrum as a peak by
the electron neutrino rest mass above Q. Here the possibility to see the CNB with KATRIN is studied.
A detection of the CNB by KATRIN seems not to be possible at the moment. But KATRIN should be able to
determine an upper limit for the local electron neutrino density of the CNB.
\end{abstract}

\body

\section{The three Cosmic Backgrounds}\label{sec1}

The three Cosmic backgrounds
\begin{enumerate}
\item Cosmic Gravitational Background (CGB; during the Big Bang (BB)),
\item Cosmic Neutrino Background (CNB; one minute after the BB) and
\item Cosmic Microwave Background (CMB;  $\sim\ 380 000 \ $ years after the BB)
\end{enumerate}
can give information about the Universe at different times after the BB indicated above.
\vspace{0.5cm}
\newline
The inflationary expansion of the Universe by about a factor $10^{26}$ between  roughly  $10^{-35}$
to $10^{-33}$  [sec] after the BB couples according to the General Relativity to
gravitational waves, which decouple after this time and their fluctuations are the seed
for Galaxy Clusters and even Galaxies. These decoupled gravitational waves run since then
with only very minor distortions through the Universe and contain a memory to the BB. The
eLISA project (Evolved Laser Interferometer Space Antenna) of the European Space Agency\cite{Pau}
with three satellites may perhaps be able to see the Cosmic Gravitational
Background (CGB). Recently the BICEP2 collaboration\cite{BICEP2} claimed to have seen
in the fluctuations and the polarization of the CMB fingerprints of the Gravitational
Wave Background originating from the Inflationary Expansion during the BB. But this was
probably a to wishful interpretation of the data.
\vspace{0.5cm}
\newline
The Cosmic Neutrino Background (CNB, often also called `relic neutrinos') decouples from
matter about 1 second after the BB in the radiation dominated era at a temperature of
$10^{10}$ Kelvin $\equiv$  1 MeV. Today due to the expansion and cooling of the
Universe the relic neutrino temperature is 1.95 Kelvin and the average neutrino density
in the Universe is 340 per  $ cm^3$ or 56 electron  neutrinos  per cm$^3$. Recently
Follin et al. \cite{Follin} interpreted data about damping of acoustic oscillations of
the Cosmic Microwave Background as an interaction with the Cosmic Neutrino Background (CNB).
\vspace{0.5 cm}
\newline
The Cosmic Microwave Background (CMB) originates about 380 000 years after the BB at a
temperature of around 3000 Kelvin and is well studied. It was detected by Penzias and
Wilson\cite{Penzias} in 1964 (Nobel Prize 1978) as byproduct of their search
for possible perturbations of the communication with satellites. Later on the CMB was
confirmed and detailed by satellite observations\cite{COBE,WMAP,Planck}.
The frequency distribution follows exactly Planck's black body formula and yields a
surprisingly identical temperature up to four digits independent of the direction
(T$_{0\gamma}$ = 2.7255(6)  Kelvin).
\begin{equation}
\epsilon(f) df =  \frac{8 \pi h}{c^3} \cdot \frac{f^3 df}{ exp(h f/k_B T_0) - 1} \:  {\rm [Energy/Volume]}
 \label{Pl}
\end{equation}

This work discusses the possibility of the detection of the Cosmic Neutrino Background
(CNB) with the KArlsruhe TRItium Neutrino (KATRIN) experiment. This work is based on our
publications \cite{Faessler4}, the publication of Drexlin et al. \cite{Drexlin}, the KATRIN
Design report \cite{Design} and numerous discussions with Drexlin and Weinheimer \cite{Drexlin2}.
\newline \\
Different methods have been discussed in the literature to search for these relic
neutrinos \cite{Ringwald, Wigmans, Ringwald2, Weiler, Vega, Munyaneza, Paes1}.

The search for the CNB with the induced beta decay \cite{Faessler4,Lazauskas} as
originally proposed by Steven Weinberg \cite{Weinberg} seems to be the most promising
using the reaction:
\begin{equation}
 \nu_e + {^3}{\rm H}  \rightarrow  {^3}{\rm He} + e^-.
\label{induced}
\end{equation}
The signal would show up by a peak in the electron spectrum with an energy of the
neutrino mass above the Q value.

\section{The Cosmic Photon or Microwave Background (CMB)} \label{sec2}

The Cosmic Background Radiation (CMB) was detected in 1964 by Penzias and Wilson\cite{Penzias} .
The frequency distribution follows very accurately Planck's
black body radiation law (\ref{Pl}). As already mentioned the temperature
parameter T$_0$ fitted by (\ref{Pl}) is independent of the observation
direction up to the fourth digit.
The average temperature of the photons in the CMB is about 3 T$_0$. Today
the temperature of the CMB is:
\begin{equation}
{\rm T}_{0,rad} \equiv {\rm T}_{0,\gamma} = 2.7255 \pm 0.0006  \: {\rm [Kelvin]}.
\label{Temp}
\end{equation}

The Stefan-Boltzmann law for the energy density originates by integration over
all frequencies $ \it f $:
\begin{equation}
\epsilon_{rad}  =  \varrho_{rad}c^2 = \alpha T_0^4.
\label{SB1}
\end{equation}

As soon as the electrons get permanently bound to the protons and the Helium nuclei
the universe is electrically neutral and since then the photons move freely as the CMB.

\section{When decouple the Neutrinos from Matter?} \label{sec3}

Since the neutrinos interact with matter only by the weak interaction they decouple
much earlier than the photons, which interact by the electromagnetic force. The
decoupling of the neutrinos from matter is determined by the competition of the
expansion rate of the Universe (Hubble parameter) and the interaction rate of the
neutrinos with matter. The Hubble parameter using the Friedmann equation \cite{Friedmann, Liddle}
and the Stefan-Boltzmann law (\ref{SB1}) is given by:
\begin{equation}
 H = \frac{\dot{a}}{a} = \sqrt{\frac{8 \pi G}{3} \varrho_{\rm total}} =
\sqrt{\frac{8 \pi \varrho}{3 M_{\rm Planck}^2}} \propto T_0^2 \propto a^{-2}.
\label{Hubble}
\end{equation}
The Universe expansion rate is given by the Hubble parameter. It decreases with
the square of the temperature. The variable `a' is any length scale describing
the expansion of the Universe. The neutrino reaction rate reduces with the fifth
power $T^5$  of the decreasing temperature:
\begin{equation}
  \Gamma = n_\nu<\sigma v> ~\approx~
  T_0^3~ G_{\rm Fermi}~ T_0^2 ~=~ G_F~ T_0^5~ \propto~ a^{-5}; \hspace{0.5cm} v \approx c=1 .
\label{neutrinoR}
\end{equation}

When the Hubble expansion rate (\ref{Hubble}) is getting faster or about
equal to the neutrino reaction rate (\ref{neutrinoR}), the neutrinos decouple
from matter.
\begin{equation}
 \frac{\Gamma}{H} \doteq 1 = \bigg( \frac{k_B \ T_0}{1\  {\rm MeV}}\bigg)^3.
\label{Decup}
\end{equation}
Thus the decoupling temperature corresponds to: $T_0$ = 1 MeV $\equiv$ $10^{10}$[Kelvin].
Today the temperature of the neutrinos is 1.95 Kelvin.

Since in the radiation dominated era one has
\begin{eqnarray}
 T_0  \: = \bigg(\frac {45 \hbar^3 c^5}
{32\pi^3 G}\bigg)^{1/4}~ \frac{1}{ t^{1/2}} \equiv  1.3\bigg(\frac{t}{sec} \bigg)^{-1/2} \ [MeV].
\label{Decoupling}
\end{eqnarray}
\vspace{0.3cm}
Thus the decoupling happens about\  1\ sec\ \mbox after the BB.

\section{Can KATRIN detect the Cosmic Neutrino Background (CNB)?} \label{sec4}

Fermi's Golden Rule gives for the Tritium beta-decay probability:

\begin{eqnarray}
  &&\Gamma_{\rm decay}^\beta({^3}{\rm H}) = \\
  &&\frac{1}{2 \pi^3}~ \sum \int |<{^3}{\rm He}|{\rm T}|{^3}{\rm H}>|^2 ~ 2\pi \delta(E_\nu+E_e + E_f - E_i)
 \frac{d\vec{p}_e}{2\pi^3}~ \frac{d\vec{p}_\nu}{2\pi^3};\nonumber
\label{tritium}
\end{eqnarray}

\begin{equation}
  {\rm Theory:}~~T^\beta_{1/2} = \frac{\ln2}{\Gamma_{\rm decay}^\beta({^3}{\rm H})} = 12.32~~
  {\rm yrs};
~~~{\rm Experiment:}~~~ T^\beta_{1/2} = 12.33 ~~{\rm yrs}.
\label{Thalf}
\end{equation}

The $\beta$-decay spectrum of the electron from the Tritium $\beta$-decay has the form:
\begin{eqnarray}
  \frac{dN_e}{dE} &=& K ~F(E,Z)~ p_e E_e ~ (E_0 - E_e) ~\times \nonumber\\
 && \sum^3_{j=1}|U_{ej}|^2 ~ \sqrt{(E_0 - E_e)^2 -m_{\nu j}^2} \theta{(E_0 - E_e - m_{\nu j})}
\label{decay}
\end{eqnarray}
with:
\begin{eqnarray}
&&  {\rm K= const};~~~ Q = 18.562~ {\rm keV}; ~~~ E_0 = Q + m_e; ~~~ E_e = \sqrt{m_e^2 + p_e^2};\nonumber\\
&&  E=T_e= E_e - m_e; ~~~\nu_e = \sum_{j=1}^3 U_{ej}  \nu_j.
\label{def}
\end{eqnarray}
$\nu_e$ \ flavor eigenstate; \quad $\nu_j$ \ mass eigenstate. The current upper limit on neutrino mass from
tritium $\beta$-decay experiments holds in degenerate neutrino mass region
($m_{\nu 1}\simeq m_{\nu 2} \simeq  m_{\nu 3} \simeq m_{\nu e}$ with $m_{\nu e} = \sum_{j=1}^3 m_{\nu j}/3$).    
In the Kurie plot
the Tritium decay spectrum is for a massless neutrino a straight line as a function
of the electron energy. The line hits the abscissa with the electron energy at
the Q-value. The Kurie plot is obtained by dividing (\ref{decay}) by product
$(K ~F(E,Z) ~ p_e ~ E_e)$ and taking the square root. The neutrino
mass modifies the Kurie plot in the interval $<Q-m_{\nu e}~ c^2|Q = 18.562~ {\rm keV}>$.
The capture of the relic neutrinos from the CNB (\ref{induced}) should show
as a peak in the electron spectrum at $Q + m_{\nu e}~ c^2$. The capture
probability requires the same Fermi and Gamow-Teller matrix elements as for
the decay. Thus the theoretical prediction for the capture probability should
be as accurate as the one for the decay.

\begin{eqnarray}
\Gamma_{\rm capture}^\beta (^3H)  &=&  \frac{1}{\pi} (G_F \cos{\vartheta_C})^2 F_0(Z+1, T_e)
      [B_F(^3H) + B_{GT}(^3H)] p_e T_e \times\nonumber\\
&&~~~ ~\langle n_{\nu,e}\rangle \frac{n_{\nu,e}}{\langle n_{\nu,e}\rangle}  \nonumber\\
&=&  4.2~ 10^{-25} \frac{n_{\nu,e}}{\langle n_{\nu,e}\rangle}[\mbox{\rm for one Tritium atom/year}]; \\
&&	{\rm with:} \hspace{0.3cm}
\langle n_{\nu,e}\rangle \:  = 56 \hspace{0.2cm} {\rm cm}^{-3}. \nonumber
\label{RateN}
\end{eqnarray}

The values for this effective strength of the Tritium source \cite{Kaboth} and
\cite{Faessler4} have been reduced step by step. The correct value given by
Drexlin \cite{Drexlin2} is $ 20 \: \mu g $. This means $ 2\times 10^{18} \quad  {\rm Tritium}_2 $
molecules. The capture rate of relic neutrinos is then:

\begin{equation}
\mbox{Capture rate at KATRIN: } N_\nu ({\rm KATRIN}) = 1.7~ 10^{-6}~ \frac{n_{\nu,e}}{\langle n_{\nu,e}\rangle}.
\label{RateK}
\end{equation}

If one uses the average relic neutrino number density $\langle n_{\nu,e}\rangle \: = \: 56 \: {\rm cm}^{-3} $,
one predict only every 590 000 years a count. But there is the hope, that the local relic
neutrino density in a galaxy increases by gravitational clustering. Ringwald and Wong
\cite{Ringwald} calculated, that relic neutrinos can cluster on the scale of
a single galaxy and their halo  and if one uses the proportionality to the baryon
overdensity (see Lazauskas et al. \cite{Lazauskas}), then one can expect very
optimistically an overdensities up to a factor $ n_{\nu,e}/\langle n_{\nu,e}\rangle \: \le  10^6 $
in our neighbourhood. With this very optimistic overdensity for the relic neutrinos
of $ 10^6$ one obtains from equation (\ref{RateK}):

\begin{equation}
  N_\nu ({\rm KATRIN}) = 1.7~ 10^{-6}~ \frac{n_{\nu,e}}{\langle n_{\nu,e}\rangle} \: ~{\rm [year^{-1}]}
  \approx 1.7 \: ~{\rm [counts\:  per\: year]}.
\label{RateKK}
\end{equation}

This seems not possible to measure for the moment. One way out would be to increase
the effective activity of the tritium source. An effective mass of 2 milligrams
Tritium would mean with the above optimistic estimate of the relic neutrino number
overdensity $ n_{\nu,e}/\langle n_{\nu,e}\rangle \:  \approx  10^6 $ about $170$ counts per year,
which should be feasible. But it should be possible with KATRIN, in its present
form, to determine an upper limit for the local relic neutrino density of the CNB. \\[0.5cm]
Is it possible to increase the Tritium source strength by factor 100? Increasing the
source strength one has to keep in mind the following requirements.

\vspace{0.6 cm}
{\bf \large Requirements:}
\vspace{0.2 cm}
\begin{enumerate}
\item {\bf Maximize detectable 18.6 keV electrons}, which contain the information on
the neutrino mass and the capture of the relic neutrinos.
\item { \bf 1 eV energy resolution of the spectrometer.}
\item { \bf Conserve the orbital angular momentum of the electrons} for the cyclotron
motion along the magnetic field lines from the source to the detector.
\item { \bf Conserve the magnetic flux.}
\item { \bf Focus a large number of 18.6 keV electrons into the 9 cm wide transport
channel} to the spectrometer.
\end{enumerate}
\vspace{0.5cm}
{\large \bf Discussion of these Requirements:}
\\[0.3cm]
(i) { \bf The increase of the Tritium source strength} is limited by the scattering of
the emitted electrons by the tritium gas \cite{Design}. After the mean free path

\begin{equation}
\lambda_{\rm mean\  free \ path} = \frac{1}{\rho ~ \sigma({\rm electron-tritium})} = d_{\rm free}
\label{mean}
\end{equation}

only about $ 37 \% $ decay electrons have not yet scattered. All the others including
also electrons with the maximum energy, which contain the information on the neutrino
mass and on the relic neutrino capture by Tritium, are lost for the measurement
\cite{Design}. Thus an increase of the Tritium source strength beyond the 20 micrograms
reduces the number of counts of 18.6 keV electrons and is counter productive.
\vspace{0.3cm}
\newline

(ii) If one request for the spectrometer an { \bf energy resolution of 1 eV}, one can
afford to have only 1 eV of energy in the spectrometer in the perpendicular cyclotron
motion. The spectrometer can by an opposing electric field of around 18 600 Volt control
the longitudinal energy of the electron motion along the magnetic field lines and allow
only electron to pass with  energies around Q = 18.562 keV. The information on the neutrino
mass and on the capture of the neutrinos from the CNB  lies close to the upper end point
of the electron spectrum in the interval $ < Q - a~ m_{\nu e}~ c^2| Q + a~ m_{\nu e}~ c^2>$
with a number `a' roughly between 2 and 20.

\begin{figure}[ht]
\centerline{
\includegraphics[width=4in]{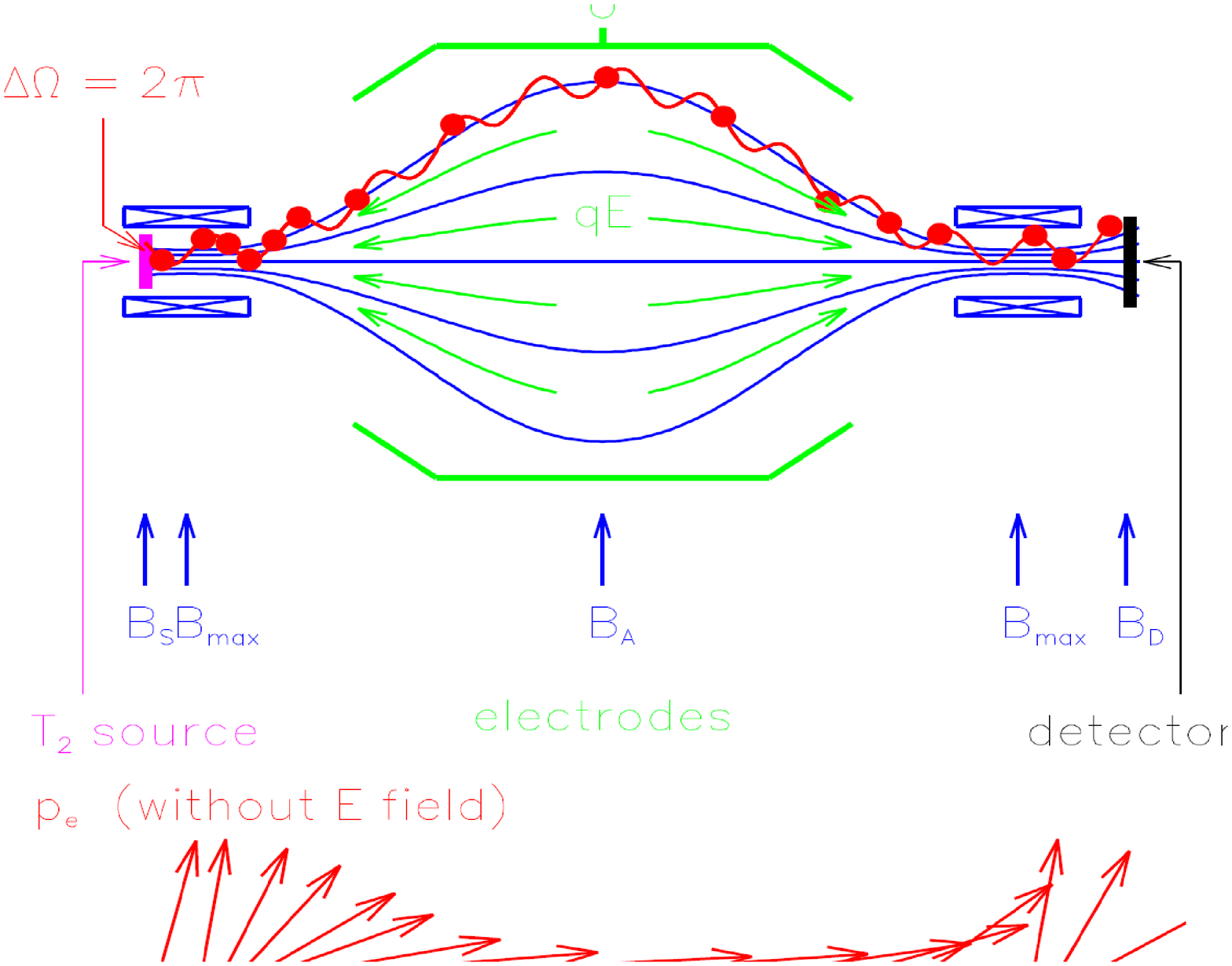}
}
\caption{Magnetic fields in the spectrometer. The magnetic field at the Tritium $T_2$
source $B_S$  focuses ideally half of the the electrons in forward direction. In reality
$B_{\rm max}$ contracts only part of the forward electrons into the transport channel with
about 9 cm diameter. A to strong increase of the field $B_{\rm max}$ yields a magnetic mirror
and reflects all electrons. The electrons follow in cyclotron rotation the magnetic field
lines. Angular momentum conservation requests, that $ E_{\perp}/B$ is constant along the
electron trajectory. Thus a small $E_{\perp}$ = 0.93 eV in the middle of the spectrometer
requires there a weak magnetic field of about 3 Gauss. Energy conservation then takes care,
that the electron momentum in the middle of the spectrometer points forward into the
longitudinal direction. The size and direction of the electron momentum is indicated at
the lower end of the figure. This an electric opposing field (not included in this figure)
allows only electrons around the Q value containing the information about the neutrino
mass and also about the Cosmic Neutrino Background (CNB) to reach the detector on the right.
the electron momenta, direction and size, are indicated at the lower part of the figure.
Figure by Christian Weinheimer, Univ. Muenster, Germany, reproduced by his permission.
\label{Spectrometer}}
\end{figure}

(iii)
{ \bf The conservation of the angular momentum}  in the circular cyclotron motion
of an electron and also the corresponding magnetic moment of this ring current must be
conserved along the trajectory of the electron from  the source to the detector.
This requires a fixed ratio of the perpendicular energy $E_{\perp}$ over the local
magnetic field along the longitudinal orbit of the electrons.

\begin{equation}
 |\vec{L}| = |\vec{r}\times \vec{p}| \propto \mu = const \propto \frac{E_{i\perp}}{B_i} = \frac{E_{f\perp}}{B_f}
\label{moment}
\end{equation}

An energy resolution of about $ \Delta E \ = \ 0.93\  eV $ thus requires:

\begin{eqnarray}
  \Delta E = 0.93 ~ {\rm eV}\ = \ E_{f\perp}\ = \  \frac{B_f}{B_i}~ E_{i\perp} \
  = \ \frac{ 3 \ {\rm Gauss}}{360 \ {\rm Gauss}}~ E_{i\perp}
\label{resolution}
\end{eqnarray}

With $B_i = 3.6$  Tesla and of the 18.6 keV at the source 10 keV in the perpendicular
motion and a required resolution of $E_{f\perp} = 0.93$ eV one obtains for the magnetic
field in the spectrometer a small value of about $B_f = 3$ Gauss.
\vspace{0.3cm}
\newline

(iv) { \bf The magnetic flux}  at the present source is given by

\begin{equation}
{\rm Magnetic~ Flux~~(source)} = 53~ {\rm cm}^2 \times 3.6~ {\rm Tesla} \approx 190~ {\rm Tesla}~{\rm cm}^2.
\label{flux1}
\end{equation}

The flux must be the same in the spectrometer requiring a perpendicular area of
the spectrometer of 63.6 m$^2$.

\begin{equation}
{\rm Magnetic ~Flux~(spectrometer)} = 63.6~ {\rm m}^2 \times 3~ {\rm Gauss} \approx 190~  {\rm Tesla}~{\rm cm}^2.
\label{flux2}
\end{equation}

\begin{figure}[ht]
\centerline{
\includegraphics[width=4in]{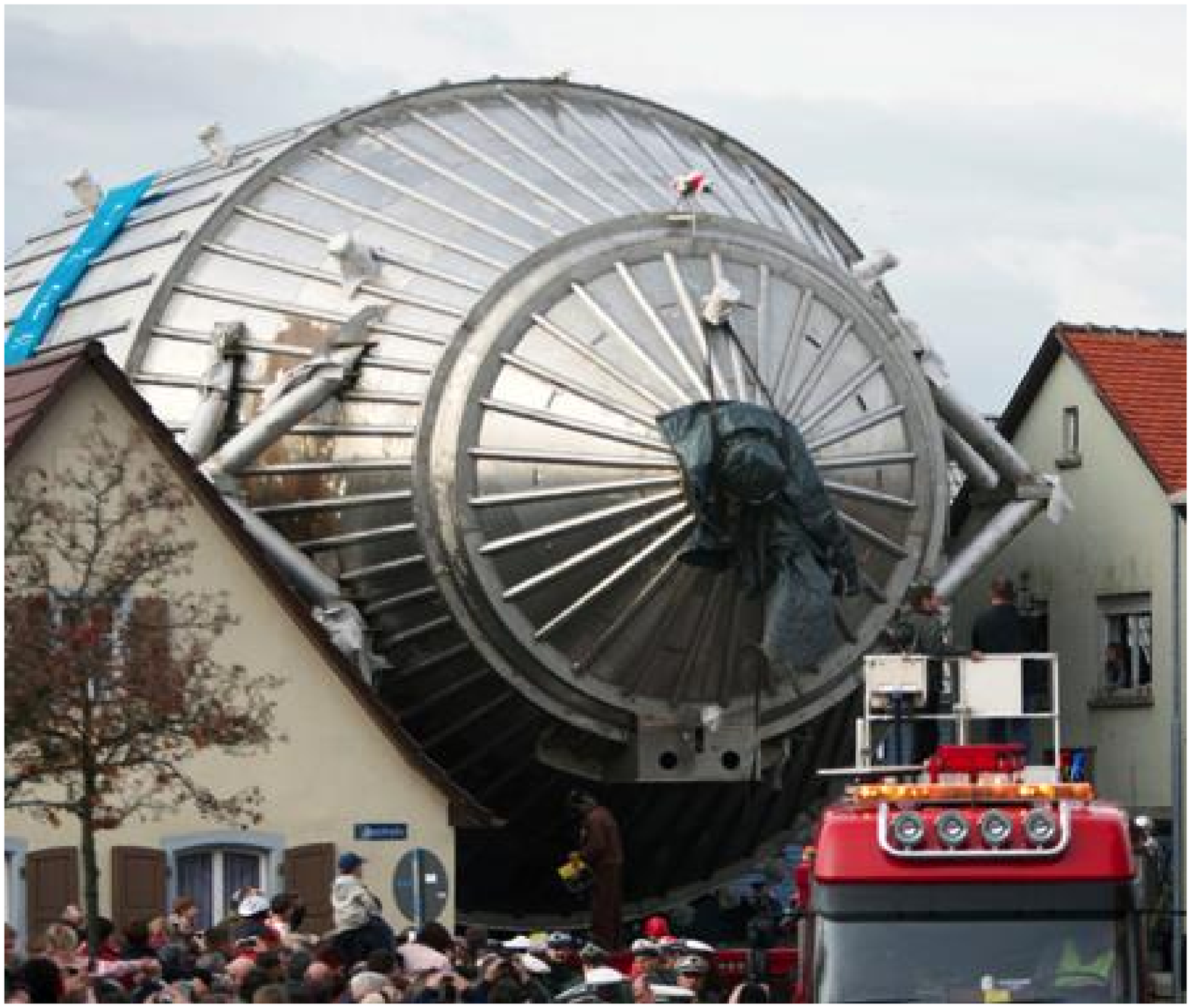}
}
\caption{KATRIN spectrometer. The conservation of the magnetic flux and energy resolution
of 0.93 eV requires a spectrometer with a diameter of about 10 meters. The figure is
reproduced from the KATRIN collaboration with permission of the speakers Guido Drexlin,
KIT Karlsruhe and Christian Weinheimer, Univ. Muenster, Germany.
\label{KATRIN}}
\end{figure}

The perpendicular area of $63.6~{\rm m}^2$ of the spectrometer required to conserve
the magnetic flux, yields for an energy resolution of 0.93 eV a large size of
the spectrometer with about 10 meter diameter. See figure (\ref{KATRIN}). \\[0.5cm]
The low magnetic field in the spectrometer of only 3 Gauss is needed to
transform the electron momenta, which are at the source almost perpendicular
to the beam direction due to the cyclotron motion, into almost a translational
direction (see figure 9 of the KATRIN design report \cite{Design} and figure
\ref{Spectrometer} of this contribution), to reject with an electric opposing
field all electrons almost up to the tritium Q-value ($Q = 18.562$  keV ).
To increase the source strength by a factor 100  with a corresponding increase
of the source area to $ 5000 \ {\rm cm}^2, ~80\ {\rm cm} $  diameter, one needs also to increase
the spectrometer cross section by a factor 100 or the diameter to about 90 m,
which is not possible.
\vspace{0.3cm}
\newline

(v) { \bf To focus a large number of 18.6 keV electrons at the source into the
9 cm wide transport channel} one needs the large magnetic field of 3.6 Tesla.
This then requires by magnetic flux conservation and the requirement of 1 eV energy
resolution the large size of the spectrometer.
\newline
With the resolution of $\Delta E \ = \ 0.93$ eV  KATRIN hopes to reduce the
upper limit of the electron neutrino mass to about 0.2 eV   $(90 \% \  {\rm C.  L.} )$.
Fitting at the upper end of the Kurie plot at $Q - m_{\nu,e}$ of the electron
spectrum the KATRIN collaboration hopes to determine the Q-value and the neutrino mass.
The electron peak due to the capture of the relic neutrinos lies at $ Q +m_{\nu,e} $ .
The neutrino mass and the energy resolution and the background remain the same as
for the determination of the neutrino mass. One has only one additional fit parameter
more (or two, if one counts the width of this peak), the counts in the peak at $Q + m_{\nu,e} $.
At the moment it does not seem possible to detect with a KATRIN type spectrometer
the Cosmic Neutrino Background. But one should be able to give an upper limit for
the local relic neutrino overdensity $ n_{\nu,e}/\langle n_{\nu,e}\rangle $ in our Galaxy.

\section{Alternative Approaches}  \label{sec5}

Two methods are discussed in the literature to go beyond KATRIN for the determination
of the electron antineutrino mass \cite{Kaboth} and the detection of the Cosmic Neutrino
Background (CNB) \cite{Ptolemy}.\\
The `Project 8' at Santa Barbara and MIT wants to measure the electron antineutrino mass
by the radiation emitted from the Tritium beta decay electrons moving in cyclotron
resonances (rotations) along a longitudinal magnetic field \cite{Kaboth}. This requires
special antennas along the electron trajectory from the source to the detector and a
detailed analysis of the radiation spectrum. The opinion of the experts of the feasibility
of this approach to determine the neutrino mass varies.  \\
The PTOLEMY project (Princeton Tritium Observatory for Light, Early-Universe, Massive-Neutrino Yield)\cite{Ptolemy} of the Princeton Plasma Physics Laboratory wants to measure neutrinos from the CNB.  The planned Tritium source is exceptionally intense with a total mass of 100 grams.  A single graphene layer backed by a substrate can bind Tritium by sub-eV energies at each Carbon intersection on the surface directly facing vacuum.  Maximal hydrogenation of the graphene surface achieves $3~ 10^{15}$ Tritium atoms per cm$^2$.  Since the probability to scatter for one 18.6 keV electron on $^{12}$C  in one graphene layer is about 3.6 percent, one is limited to two or three stacked graphene layers (see ref.\cite{Ptolemy} page
two).  This limits the weight of Tritium to about 1 microgram per cm$^2 $ with two layers.  If one assumes two layers with 1 microgram per $cm^2$ one needs for 100 grams of Tritium roughly $10^8$  cm$^2$  =  10 000 m$^2$ of double graphene layers with a free way from the source to the detector.  The PTOLEMY proposes a time-dependent system incorporating the Project 8 antenna technology to significantly reduce the phase space requirements, in additional to alternative distributed geometries for the MAC-E filter inspired by KATRIN.  Although the PTOLEMY collaboration discusses these problems in detail \cite{Ptolemy}, it is obvious that such a source is extremely difficult to realize.

\vspace{0.5cm}

\section{Conclusion}

\begin{itemize}

\item The average relic electron neutrino number density of $\langle n_{\nu,e}\rangle$ = 56 cm$^{-3}$ KATRIN
yields only every 590 000 years a count from the CNB with KATRIN. But the local overdensity
due to gravitational clustering of the neutrinos in our galaxy will increase the
counting rate. Estimates for this overdensity $n_{\nu,e}/\langle n_{\nu,e}\rangle $  vary widely
from about $10^2$ to $10^6$ and depend on the mass of the relic neutrinos. The very
optimistic value of the local overdensity of $ 10^6$ yields with KATRIN 1.7 counts
per year. Increasing  the effective mass of the Tritium source from 20 micrograms
to 2 milligrams would lead to 170 counts per year. Detection would be possible.

\item The second problem could be the energy resolution of the KATRIN spectrometer
of $\Delta E$ = 0.93 eV. With this resolution KATRIN expects to extract from
the electron spectrum at the upper end of the Kurie plot at $ Q - m_{\nu,e} $ a very
accurate Q-value with an error of milli-eV or less and an upper limit of the electron
antineutrino mass of about $ m_{\nu,e} \le$  0.2 [eV]  90 $\%$  C.L..
If one can fit the Q value and the electron neutrino mass accurately enough, the
position of the electron peak from the induced capture of the relic neutrinos is known
to be at an electron energy of $ Q~ + ~ m_{\nu,e}~ c^2 $. The energy resolution
and the background is the same at $ Q - m_{\nu,e}~ c^2 $ and at
$ Q + m_{\nu,e}~ c^2 $. One should with KATRIN at least be able to determine an
upper limit for the local relic neutrino density.

\end{itemize}

\section*{Acknowledgements}

We thank Christian Weinheimer and Guido Drexlin for discussions about the KATRIN
spectrometer, Christopher Tully for information about PTOLEMY and the Ministry 
of Education and Science of the Russian Federation
(contract 12.741.12.0150). F. S. acknowledges the support by the VEGA Grant agency
of the Slovak Republic under the contract No. 1/0922/16 and by the Ministry of
Education, Youth and Sports of the Czech Republic under contract LM2011027. S. K.
thanks for support by the FONDECYT grant 1100582 and the Centro Cientifico-Tecnol\'{o}gico
de Valpara\'{i}so PBCT ACT-028.


\end{document}